\documentclass[pra,amsmath,amssymb,showpacs,superscriptaddress]{revtex4}
\usepackage{graphicx,epstopdf,bm}

\newcommand{\Fkt}[1]{\,\mathsf {#1}}
\ifx\Tr\renewcommand{\Tr}{\Fkt{Tr}}
\else\newcommand{\Tr}{\Fkt{Tr}}
\fi
\newcommand{\bra}[1]{\ensuremath{\langle#1|}}
\newcommand{\ket}[1]{\ensuremath{|#1\rangle}}
\newcommand{\braket}[1]{\ensuremath{\langle#1\rangle}}

\begin{document}

\title{Optimal qudit operator bases for efficient characterization of
  quantum gates} 

\author{Daniel M. Reich}
\affiliation{Theoretische Physik, Universit\"{a}t Kassel,
  Heinrich-Plett-Str. 40, D-34132 Kassel, Germany} 

\author{Giulia Gualdi}
\affiliation{Dipartimento di Fisica ed Astronomia, Universit\`a di
  Firenze, Via Sansone 1, 50019 Sesto Fiorentino, Italy}
\affiliation{QSTAR, Largo Enrico Fermi 2, 50125 Firenze, Italy}

\author{Christiane P. Koch}
\affiliation{Theoretische Physik, Universit\"{a}t Kassel,
  Heinrich-Plett-Str. 40, D-34132 Kassel, Germany} 
\email{christiane.koch@uni-kassel.de}

\begin{abstract}
  For certain quantum operations acting on qubits, there exist bases of 
  measurement operators such that 
  estimating the average fidelity becomes efficient.
  The number of experiments required is then
  independent of system size and the classical computational resources
  scale only polynomially in the number of qubits. Here we
  address the question of how to optimally choose the measurement
  basis for efficient gate characterization when  replacing two-level
  qubits by $d$-level qudits. We define optimality in terms of the maximal
  number of unitaries that can be efficiently characterized. Our
  definition allows us to construct the optimal measurement basis in
  terms of their spectra and eigenbases: The measurement operators are
  unitaries with $d$-nary spectrum and partition into $d+1$ Abelian
  groups whose eigenbases are mutually unbiased. 
\end{abstract}

\pacs{03.65.Wj,03.67.Ac}

\date{\today}
\maketitle

\section{Introduction}
\label{sec:intro}

The development and maintenance of quantum devices requires the
capability to verify their proper functioning. This is quantified by 
suitable performance measures such as the average gate
fidelity~\cite{NielsenChuang}. In order to determine the gate fidelity
in a given experimental setup, no matter what is the specific
protocol, one needs to define a set, or, more precisely, a complete
and orthonormal basis, of measurement
operators~\cite{NielsenChuang}. The choice of 
measurement operators is typically dictated by considerations of
experimental convenience such as the requirement of local
measurements in the sense that each operator can be measured in a separable
eigenbasis. 

Additional considerations become important for 
certain classes of quantum operations, namely those that map a
suitable basis of measurement operators onto itself, up to a phase
factor. For qubits,  
Pauli measurements represent such an operator basis. The 
corresponding unitary operations are termed Clifford gates; they 
facilitate fault-tolerant 
computation~\cite{GottesmanChaos99} and yield a universal set when 
augmented by the proper local phasegate~\cite{Boykin00}. The property
of Clifford gates to map the operator basis onto itself, up to a phase
factor, can be
exploited to obtain protocols for determining the average gate fidelity
that require a  number of experiments that is independent of
system size and classical computational resources that scale only
polynomially in the number of information
carriers~\cite{FlammiaPRL11,daSilvaPRL11,MagesanPRL11}.   

When replacing two-level qubits by $d$-level qudits, one is faced with
the problem that the $d$-dimensional generalizations of the Pauli
measurement basis cannot be Hermitian and unitary at the same time. 
Different choices of measurement bases exist that correspond to 
different numbers of unitaries for which 
efficient characterization is possible~\cite{Giulia14}. 
This raises the question of the optimal choice for the  measurement
basis.  

Here we address this question by defining optimality in terms 
of the maximal number of unitaries that can be efficiently
characterized and use this definition to construct the optimal
measurement basis in terms of their spectra and eigenbases.
We find the optimal measurement basis to consist of 
unitaries with $d$-nary spectrum that partition into $d+1$ Abelian
groups whose eigenbases are mutually unbiased. 
Our result motivates the use of the generalized Pauli
group~\cite{GottesmanChaos99,BermejoQuantInf14} 
as an optimal measurement basis, not least because of its close
connection to mutually unbiased
bases~\cite{WoottersAnnPhys89,BandyoAlgo02,LawrencePRA04}. 

The paper is organized as follows: We first define optimality of an
operator basis for estimating the average fidelity of quantum gates in
Sec.~\ref{sec:problem}. 
In the following, we use this definition of optimality 
in Sec.~\ref{sec:spec} and~\ref{sec:ebases} to construct the 
operators that make up the optimal set in terms of their spectra and
eigenbases for the case that the Hilbert space dimension $d$ is a prime
number. The construction
will allow us to show that the optimal operator basis consists of
unitaries with $d$-nary spectrum (i.e., the spectrum is made up of the
$d$th roots of unity) and partitions into $(d+1)$ Abelian groups whose
eigenbases are mutually unbiased. The latter is demonstrated in
Sec.~\ref{sec:MUB}. 
For the case that $d$ is not prime,
we construct the measurement operators as tensor products and can thus
reuse our results obtained for $d$ prime in
Sec.~\ref{sec:tensor}. Section~\ref{sec:concl} concludes.

\section{Problem statement}
\label{sec:problem}

We consider a Hilbert space of dimension $d$ with $d$ prime. Any suitable operator
basis $\mathcal M$ defined on this Hilbert space 
must be complete and orthonormal. Unitaries that map the
operator basis onto itself, up to a phase factor, can be efficiently
characterized, for example by employing Monte Carlo estimation of the
average fidelity~\cite{FlammiaPRL11,daSilvaPRL11}.  
Correspondingly, we define the set of unitaries 
$\mathcal{U}_{\mathcal{M}}$ by the property that for all
$U\in\mathcal{U}_{M}$ and $M_{i}\in\mathcal{M}$ there exists a
$M_{j}\in\mathcal{M}$ such that $UM_{i}U^{\dagger}=e^{i\phi_{i}}M_{j}$
with $\phi_{i}\in\mathbb{R}$ some phase. This property guarantees a
relevance distribution for the Monte Carlo sampling with 
$d^{2}$ non-vanishing entries which is the minimal amount~\cite{Giulia14}.
Furthermore, these entries all have equal magnitude.

We define an operator basis set $\mathcal{M}$ to be optimal, 
$\mathcal M^\star$, if 
$\left|\mathcal{U}_{\mathcal{M}}\right|=u_{max}$
where
$u_{max}=\text{max}_{\mathcal{M}'}\left|\mathcal{U}_{\mathcal{M}'}\right|$
and $\left|\cdot\right|$ denotes the cardinality of a set. 
That is to say that an operator basis 
$\mathcal{M}$ is optimal if the number of unitaries that map 
the basis onto itself is maximal amongst all possible operator bases.
The map here is to be understood as the conjugation $U:M\mapsto UMU^{\dagger}$.

\section{Spectral properties}
\label{sec:spec}

Completeness of the operator basis implies that the set $\mathcal M$ contains
$d^2$ elements. We include the identity in $\mathcal M$ since $\openone$ is
mapped onto itself by \textit{all} unitaries. This provides a good starting point
for the construction of $\mathcal M^\star$ which requires all $M_i$ to
be mapped to some $M_j\in \mathcal M^\star$ by as \textit{many}
unitaries as possible.  We can thus restrict the following
discussion to the $d^{2}-1$ traceless operators in $\mathcal{M}$. 
Tracelessness of the remaining operators $M_1, M_2, \ldots,M_{d^2-1}$
in $\mathcal{M}$ follows from their
orthogonality to the identity. We denote this set by $\tilde{\mathcal
  M}$, i.e., $\tilde{\mathcal M}=\mathcal M\setminus\openone$.

By assumption, $M_1\in \tilde{\mathcal M}$ is mapped  to some $M_j\in
\tilde{\mathcal M}$ for any unitary 
$U\in \mathcal U_{\mathcal  M}$,  i.e., 
$U M_{1}U^{\dagger}=e^{i\phi_{1}}M_{j}$ with  $\phi_{1}$
a phase. $M_j$ can either be $M_1$ itself, and we speak of a
cycle of degree 1, or some other element of  $\tilde{\mathcal M}$. In
the latter case, we take $j=2$ without loss of generality. Applying
the map to $M_2$, $UM_{2}U^{\dagger}=UUM_{1}U^{\dagger}U^{\dagger}
=U^{2}M_{1}\left(U^{\dagger}\right)^{2}$
yields either a result proportional to $M_1$, in which case we have a cycle of degree 2, or
a result proportional to another $M_j$ for which we can set $j=3$. Note that the outcome of 
$UM_{2}U^{\dagger}$ cannot be $M_2$ if $M_2=UM_{1}U^{\dagger}$ due to the bijectivity of
rotations. The cycle will necessarily be closed after a number of 
repeated applications of the map since this always leads to an element of
$\tilde{\mathcal M}$, and there are only $d^2-1$ elements in
$\tilde{\mathcal M}$. We define the cycle to be of degree $n$ on the set
$\tilde{\mathcal{M}}$ if $U^{n}M_{1}\left(U^{\dagger}\right)^{n}=e^{i\phi_{n}}M_{1}$
with $n\leq d^{2}-1$ and $\phi_{n}$ a phase. 

An iterative argument shows that every operator $M$ in the set
$\tilde{\mathcal M}$ is contained in at least one cycle. To see this, choose
the lowest $i$ such that $M_{i}$ is not contained in a previously
considered cycle and apply $U$ repeatedly on $M_{i}$ until
$U^{n}M_{i}\left(U^{\dagger}\right)^{n}=e^{i\phi_{n}}M_{i}$. This
procedure can be repeated until the complete set $\tilde{\mathcal M}$
is exhausted. In fact, for a specific $U \in\mathcal U_{\mathcal M}$,
every operator $M\in\tilde{\mathcal M}$ appears exactly once in all
the cycles generated by this $U$.
As a consequence, the sum over the degrees of all cycles generated by
$U$ needs to be $d^{2}-1$. This can be seen follows: Since
rotations are bijective, $U:M_{i}\mapsto
UM_{i}U^{\dagger}=e^{i\phi_{i}}M_{j}$ induces a mapping between the
integers $i$ and $j$ which is also bijective. Therefore each $i$ can
also only occur in one cycle. The degree of a cycle measures how many
indices $i$ are present in this cycle. Since the total number of
indices is $d^2-1$, summing over the degrees of all cycles must amount
to $d^{2}-1$. The two extreme cases are that there are $d^2-1$ cycles
of degree 1 (e.g. when $U$ is the identity) or that there is one cycle
of degree $d^2-1$. 

For the operator basis to optimal, the unitary mappings on $\tilde{\mathcal M}$
should allow for arbitary cycle structures, i.e., cycles
of degree $1$, a single cycles of degree $d^{2}-1$, and anything
in between. This guarantees that the number of unitaries in $\mathcal
U_{\mathcal M}$ is not limited by the cycle structure. Specifically,
for a cycle of degree $d^2-1$ to exist, all operators in the
set $\tilde{\mathcal{M}}$ must have the same 
spectrum~\footnote{There is always the freedom
  of a global phase on the spectrum of each measurement operator. It does
  not influence the relevance distribution and thus does not affect
  the property of 
  efficient characterizability in any Monte Carlo protocol. For this
  reason we set the global phase to zero. Our term
  'the same spectrum' therefore corresponds to, strictly speaking, 
  'the same spectrum up to a global phase'.}. This is due to all
elements in this cycle emerging from one another by unitary
transformation which leaves the spectrum invariant. The requirement of
an identical spectrum for all $M_i\in\tilde{\mathcal M}^\star$ automatically
also allows for the existence of cycles of all other degrees. We
denote the spectrum of the operators in the set $\tilde{\mathcal{M}}$ by
$\text{spec}\left(\tilde{\mathcal{M}}\right)$. 

The condition of an identical spectrum together with the property that
the operator basis is mapped onto itself by $U\in \mathcal U_{\mathcal{M}}$
implies that the eigenvalues must form a closed cycle: From
$UM_{i}U^{\dagger}=e^{i\phi_{i}}M_{j}$, we obtain for the spectrum
$e^{i\phi_{i}}\text{spec}\left(M_{j}\right)\overset{!}{=}\text{spec}\left(M_{j}\right)$,
i.e., if $\lambda\in\text{spec}\left(\tilde{\mathcal{M}}^\star\right)$ then
$e^{i\phi_{i}}\lambda\in\text{spec}\left(\tilde{\mathcal{M}}^\star\right)$.
Multiplication by a complex number $e^{i\phi_{i}}$ corresponds
to rotating the eigenvalue by an angle $\phi_{i}$ in the complex
plane. Unless $\phi_{i}$ is a multiple of $2\pi$, a new eigenvalue 
$\mu=e^{i\phi_{i}}\lambda$ is obtained.
Each application of $U$ thus rotates an eigenvalue onto the next one
until the cycle is closed. The degree of the cycle on the eigenvalues
can be at most $d$ since the operators in $\tilde{\mathcal M}$ can at most
have $d$ distinct eigenvalues. Similarly to asking above for the
existence of operator cycles of all degrees, asking for the longest
eigenvalue cycle ensures that the number of unitaries in $\mathcal
U_{\mathcal{M}}$ is not unnecessarily restricted. This implies
$\left(e^{i\phi_{i}}\right)^{d}=1$, i.e., the smallest possible
rotation angle between two distinct eigenvalues is
$\phi_{i}=\frac{2\pi}{d}$. As a consequence the spectrum in polar representation
$\lambda_{i}=r_{i}e^{i\phi_{i}}$ needs to fulfill $r_{i}=r=\text{const.}$
and $\phi_{i}=\frac{2\pi k}{d}+\phi_{0}$ with $\phi_{0}$ arbitrary
such that any rotation by $\frac{2\pi}{d}$ leaves the spectrum invariant.
The normalization condition on the operator basis $\mathcal{M}$ yields
$r=1$. Since a global phase on the spectrum is
physically irrelevant we can choose $\phi_{0}=0$.
 
To summarize, for an operator basis $\mathcal M$ not to restrict the
number of unitaries that map $\mathcal M$ onto itself, the spectrum is
identical for all $M\in \mathcal M\setminus\openone$ 
and $d$-nary, i.e., it consists of the $d$th roots of unity:
\begin{equation}
  \label{eq:spectrum}
  \text{spec}\left(\mathcal{M}^\star\right)=
  \left\{ \lambda_{k}=e^{i\frac{2\pi k}{d}} ~|~ k=0,\ldots,d-1\right\} \,.
\end{equation}
In particular, this requires all measurement operators in
$\mathcal{M}$ to be unitary. As can be seen from
Eq.~\eqref{eq:spectrum}, the operators in $\mathcal{M}^\star$ cannot be unitary and Hermitian
at the same time for $d>2$. For a discussion of
non-Hermitian, unitary measurements please see Ref.~\cite{Giulia14}
and references therein.

\section{Properties of the eigenbases}
\label{sec:ebases}

In the previous section, we have used the transformation of the
operators $M \in\mathcal M$ under a special class of rotations
together with the requirement not to restrict the number of unitaries
in this class to derive the spectral properties of the operator basis. 
We can now use orthogonality of  the operator basis,
\begin{equation}
  \label{eq:ortho}
  \Tr[M_a M_b^\dagger] = \delta_{ab} \quad\forall\; M_a,M_b\in\mathcal M\,,  
\end{equation}
to obtain information
about the eigenbases of the operators in $\mathcal{M}$~\footnote{
  To be precise, the two properties that we have not yet exploited are
  orthogonality and completeness. However, completeness immediately
  follows from orthogonality and the fact that $\mathcal M$ contains (by
  definition) $d^2$ elements.}. 
Since any orthogonal basis of the underlying Hilbert space is an
eigenbasis of the identity, i.e., the eigenbasis of $\openone$
is undetermined, we only consider the $d^{2}-1$ traceless operators in
$\tilde{\mathcal M}=\mathcal M\setminus \openone$. 

We order the eigensystem according to the complex phase in the spectrum, Eq.~\eqref{eq:spectrum},
i.e., $\lambda_{k}=e^{i\frac{2\pi k}{d}}$ for $k=0,\ldots,d-1$ and
consider two distinct arbitary measurement operators $M_{a}$ and $M_{b}$, 
$a\neq b$, with corresponding eigenbases 
$\left\{ \ket{\psi_{k}^{a}}\right\} _{k=1,\dots,d}$
and $\left\{ \ket{\psi_{k}^{b}}\right\} _{k=1,\dots,d}$. 
Employing a spectral decomposition, 
$M_a = \sum _k \lambda_k \ket{\psi_{k}^{a}}\bra{\psi_{k}^{a}}$,
and expanding the trace in Eq.~\eqref{eq:ortho} in the eigenbasis of
$M_a$, we obtain
\begin{eqnarray*}
  \Tr\left[M_{a}M_{b}^{\dagger}\right] & = & 
  \sum_{klm}\lambda_{k}\lambda_{l}^{*}\braket{\psi_{m}^{a}|\psi_{k}^{a}}
  \braket{\psi_{k}^{a}|\psi_{l}^{b}}\braket{\psi_{l}^{b}|\psi_{m}^{a}}
   =  \sum_{kl}\lambda_{k}\lambda_{l}^{*}
  \left|\braket{\psi_{k}^{a}|\psi_{l}^{b}}\right|^{2}=0\,.
\end{eqnarray*}
Inserting the ordered eigenvalues yields for the trace 
\begin{eqnarray}
  \Tr\left[M_{a}M_{b}^{\dagger}\right] & = & 
  \sum_{kl}e^{i\frac{2\pi k}{d}}e^{-i\frac{2\pi l}{d}}
  \left|\braket{\psi_{k}^{a}|\psi_{l}^{b}}\right|^{2}
  =  \sum_{kl}e^{i\frac{2\pi\left(k-l\right)}{d}}
  \left|\braket{\psi_{k}^{a}|\psi_{l}^{b}}\right|^{2}\nonumber \\
  & = & \sum_{s}e^{i\frac{2\pi s}{d}}\sum_{k}\left|
    \braket{\psi_{k\oplus s}^{a}|\psi_{k}^{b}}\right|^{2}\label{eq:eigenbases}\,,
\end{eqnarray}
where in the last step we have shifted the index $s$ to run from 0 to 
$d-1$ and $\oplus$ denotes addition modulo $d$ corresponding to the
group $Z_{d}$ on the eigenbasis indices. 
Equation~\eqref{eq:eigenbases} can be interpreted as a change of basis
between the eigenbases of $M_a$ and $M_b$, 
\begin{subequations}
  \label{eq:basis_shift}
  \begin{equation}\label{eq:Uab}
    U^{ab}=\sum_{k}\ket{\psi_{k}^{b}}\bra{\psi_{k}^{a}}\,,
  \end{equation}
  together with a right-shift by $s$ in the eigenbasis of $M_a$, 
  \begin{equation}
    S^{a}\left(s\right)=\sum_{k}\ket{\psi_{k\oplus s}^{a}}\bra{\psi_{k}^{a}}\,.
  \end{equation}
\end{subequations}    
With the definitions of Eqs.~\eqref{eq:basis_shift},
we can rewrite the orthogonality condition as
\begin{eqnarray}
  \label{eq:ortho_s}
  \Tr\left[M_{a}M_{b}^{\dagger}\right] & = & 
  \sum_{s}e^{i\frac{2\pi s}{d}}\sum_{k}\left|
    \braket{\psi_{k}^{a}|S^{a}\left(s\right)U^{ab}|\psi_{k}^{a}}\right|^{2}=0\,.
\end{eqnarray}
To derive from Eq.~\eqref{eq:ortho_s}
requirements that the operator eigenbases of operators in the optimal
set $\mathcal M^\star$ must meet, we first
assume $M_a$ and $M_b$ to commute and analyze the case of
non-commuting operators in Sec.~\ref{subsec:complete} below.

\subsection{Commuting measurement operators}
\label{subsec:commuting}

Due to Eq.~\eqref{eq:spectrum}, all operators in $\tilde{\mathcal M}$
are non-degenerate. This together with the assumption
$\left[M_{a},M_{b}\right]=0$ implies that for each index $k$ there
exists an index $l$ such that $\ket{\psi_{k}^{a}}=\ket{\psi_{l}^{b}}$
and the mapping between $k$ and $l$ is bijective. That is to say that 
the eigenbases of $M_{a}$ and $M_{b}$ are the same up to reordering
which means that certain eigenvectors can correspond to different
eigenvalues. In this case,  $U^{ab}$ as defined in Eq.~\eqref{eq:Uab} 
is a permutation operator. In the eigenbasis of $M_a$, the matrix
elements of $U^{ab}$ are either zero or one and the total number of
one's is $d$. $S^{a}\left(s\right)$ is also a permutation operator
which shifts the columns of $U^{ab}$ in this representation by $s$ to the
right. This means that for all $s$, the sum over $k$ in
Eq.~\eqref{eq:ortho_s} is a non-negative integer, 
\begin{equation}
  \label{eq:c_s}
  \sum_{k}\left|
    \braket{\psi_{k}^{a}|S^{a}\left(s\right)U^{ab}|\psi_{k}^{a}} 
  \right|^{2}= c_{s}\,.  
\end{equation}
Since $U^{ab}$ and $S^{a}\left(s\right)$ are both permutation
operators, so is their product, $P^{ab}(s)=S^{a}\left(s\right)U^{ab}$. 
Note that $S^a(s=0)=\openone$,
and $c_0$ is given by the sum over the diagonal elements squared of
$U^{ab}$. For $s=1$, all columns of $U^{ab}$ are shifted to the right
by one, i.e., the first upper diagonal of $U^{ab}$ becomes the
diagonal of $P^{ab}$, and the sum over its elements squared yields $c_1$. In
other words, each $c_s$ corresponds to the sum over the diagonal of
$P^{ab}(s)$, that is the $s$th secondary diagonal of
$U^{ab}$, and thus takes a value between $0$ and $d$. 
Due to orthogonality of the operator basis, Eq.~\eqref{eq:ortho_s},
the set of integers $\left\{ c_{s}\right\} _{s=0,\ldots,d-1}$
has to fulfill the condition
\begin{equation}
  \label{eq:unityroots}
  \sum_{s=0}^{d-1}c_{s}e^{i\frac{2\pi s}{d}}=0\,.
\end{equation}
Note that, $\sum_{s=0}^{d-1}c_s=d$
since summing over all $c_s$ corresponds to summing over all elements
squared of $P(s)$, or $U^{ab}$. 
We show in Appendix~\ref{subsec:unityroots} that for $d$ prime no
linear combination with non-negative integers $c_s$ can exist that
makes the sum go to zero except if $c_s=1$ for all $s$. 

Since $c_s$ corresponds to the sum over the $s$th secondary diagonal of
$U^{ab}$, we have thus restricted all possible matrices $U^{ab}$ for a change of
basis between the eigenbases of commuting measurement operators
$M_a,M_b\in\tilde{\mathcal{M}}$ to those that contain exactly one
entry equal to one on each 
(secondary) diagonal with all other entries being zero. 
In addition, each row and each column of $U^{ab}$ also contains
exactly one entry equal to one with all other entries being zero since
$U^{ab}$ is a permutation operator. We now show that under these
constraints there exist $d-2$ distinct permutation operators
$U^{ab}$. This implies that there are $d$ orthogonal, pairwise
commuting operators with their spectrum given by
Eq.~\eqref{eq:spectrum}: $M_a$ plus the $d-2$ operators obtained by 
applying $U^{ab}$ to $M_a$ plus identity. We first show how one can
construct $d-2$ such unitaries and then prove in a second step that
these are indeed \textit{all} unitaries that fulfill the given
constraints. 

In order to construct the $d-2$ matrices $U^{ab}$ for a change of
basis, we reorder the eigenbases of $M_a$ and $M_b$ 
such that the main diagonal always contains one as
its first entry for all $b$: $U^{ab}_{11}=1$, $U^{ab}_{ii}=0$ for
$i=2,\ldots,d$. 
This reordering does not interfere with ordering 
the eigenbases of $M_a$ and $M_b$ in terms of the eigenvalues,
Eq.~\eqref{eq:spectrum}, since $M_a$ and $M_b$ can be multiplied by 
$e^{i\frac{2\pi t}{d}}$ for some $t$ without changing the
orthogonality condition. This multiplication
performs exactly the shift in the eigenbases required to ensure
$U^{ab}_{11}=1$ for all $b$. In other words: The ordering of the
eigenvalues determines the indexing of the eigenbasis of $M_b$ while
now in addition the global phase of $M_b$ is fixed. 
Then, for $d$ prime,  
a set of $d-2$ permutation operators that have on each of their
diagonals exactly one entry equal to one with all others being zero
and $U^{ab}_{11}=1$ is given by
\begin{equation}
  \label{eq:U_construction}
  \left(U^{ab}\right)_{ik}=\delta_{k,\left(i-1\right)\cdot
    b\oplus1}\quad\mathrm{with} \quad b=2,\ldots,d-1\,.
\end{equation}
The construction that leads to Eq.~\eqref{eq:U_construction} proceeds
as follows: The first row is given by the assumption
$\left(U^{ab}\right)_{11}=1$ for all $b$. In the second row,
$\left(U^{ab}\right)_{21}$ and $\left(U^{ab}\right)_{22}$  need to be
zero due to the constraints of each column and the 
main diagonal containing exactly one entry equal to one. The smallest
$j$ for which $\left(U^{ab}\right)_{2j}$ can be non-zero is thus
$j=3$. Analogously, in the third row, the smallest entry that can be
non-zero is $j=5$ (with $j=4$ being excluded by the condition on the
first upper diagonal). This construction is similar to the movement
of a knight on a chess board: one step down, two steps to the right. 
It is continued until the last row is reached to yield the first
$U^{ab}$ (with $b$ set to 2). The second $U^{ab}$ is obtained by
choosing $j=4$ in the construction of the second row. This implies a
modified movement of the knight with one step down, $b=3$ steps to the
right. Once the right boundary on the matrix is reached, the movement is
simply continued by counting from the left, as implied by the modulo
algebra in Eq.~\eqref{eq:U_construction}.
For a $d\times d$ matrix $U^{ab}$, there are $d-2$ distinct
knight-type movements since in the construction of the second row,
$\left(U^{ab}\right)_{21}$ and $\left(U^{ab}\right)_{22}$ are always
fixed and one can choose at most $j=d$, i.e., move at most $d-1$ steps to
the right. 
As shown in Appendix~\ref{subsec:unitary}, for $d$ prime, 
the construction rule, Eq.~\eqref{eq:U_construction}, yields  proper
unitary permutation operators which have on each (secondary) diagonal
only one entry equal to one. This holds only for prime
$d$.  For non-prime $d$, the above construction leads 
to a contradiction to the unitarity constraint of each column
having exactly one entry equal to one with all others being zero.

When applied to $M_a$, the $U^{ab}$ constructed according to
Eq.~\eqref{eq:U_construction} yield $d-2$ operators $M_b$ that
are orthogonal to $M_a$. We now show that
Eq.~\eqref{eq:U_construction} represents \textit{all} the unitaries
that fulfill the constraint of having exactly one entry equal to one on
each (secondary) diagonal, i.e., there are exactly $d$ commuting
measurement operators (including identity). 
As a side result, we obtain that all $M_b$
obtained from applying the $U^{ab}$ to $M_a$ are not only orthogonal to
$M_a$ but also to each other. 

The fact that, for $d$ prime,  
\textit{all} permutation operators, that have on each of their
diagonals exactly one entry equal to one with all other entries being zero
and $U^{ab}_{11}=1$, are given by Eq.~\eqref{eq:U_construction} and
that there are thus $d-2$ such unitaries can be seen as follows: 
Since $U^{ab}$ maps the eigenvectors of $M_a$ onto the eigenvectors of
$M_b$, it also corresponds to a mapping between the eigenvalues
$\lambda^a_k$ and $\lambda^b_{k^\prime}$. The fact that we fixed
$\left(U^{ab}\right)_{11}=1$ together with Eq.~\eqref{eq:spectrum}
implies $\lambda^a_0=\lambda^b_0=1$. The other eigenvalues are
redistributed according to $\lambda_{k}^{b}=e^{i\frac{2\pi}{d}\cdot k}
\;\mapsto\;\lambda_{kb}^{a}=e^{i\frac{2\pi}{d}\cdot kb}$ where the product $kb$ is to be
understood modulo $d$. Since the eigenvalue $\lambda_{kb}^{a}$ shows
up in the spectral decomposition of the $b$th power of $M_a$,
\begin{equation}
  \label{eq:Ma_power}
\left(M_{a}\right)^{b}=\left(\sum_{k}e^{i\frac{2\pi}{d}k}
  \ket{\psi_{k}^{a}}\bra{\psi_{k}^{a}}\right)^b
=\sum_{k}e^{i\frac{2\pi}{d}kb}\ket{\psi_{k}^{a}}\bra{\psi_{k}^{a}}\,,
\end{equation}
we find 
\begin{equation}
  \label{eq:M_b}
  M_{b}=\left(M_{a}\right)^{b} \quad\mathrm{with}\quad
  b=2,\ldots,d-1\,.  
\end{equation}
Moreover, $\left(M_{a}\right)^{d}=\openone$ since $kd=1$ when
interpreted modulo $d$ for all $k$. Then all powers of $M_a$
are orthonormal since, for all $b$,
\[
\Tr\left[M_{a}\left(M_{a}^{\dagger}\right)^{b}\right]
=\Tr\left[M_{a}M_{a}^{\dagger}\left(M_{a}^{\dagger}\right)^{b-1}\right]
=\Tr\left[\left(M_{a}^{\dagger}\right)^{b-1}\right]=
\begin{cases}
  1 & \text{if}\; b\,\text{mod}\,d=1\\
  0 & \text{otherwise}
\end{cases}\,.
\]
The last step follows from the fact that $M_{a}^{b-1}$ has the same spectrum
as $M_{a}$ and is consequently traceless, unless $b-1 = d$ where we
obtain identity. This is evident from Eq.~\eqref{eq:Ma_power}. Adjungation 
of the operator just returns the complex conjugated result for the trace.
Since this result is real in either case, it is unaffected by adjungation.
Finally, the maximal number of commuting,
pairwise orthogonal unitaries $M_a$ defined on a $d$-dimensional
Hilbert space 
is $d$. This can be seen by considering their common eigenbasis
$\left\{\ket{\psi_{k}}\right\}_{k=1,\dots,d}$. Any linear combination
of the commuting, pairwise orthogonal unitaries $M_a$ also has this
eigenbasis. We can thus employ the common eigenbasis to construct a
representation of any operator $M$ with this eigenbasis, 
$M=\sum_{k=0}^{d-1}\lambda_k\ket{\psi_{k}}\bra{\psi_{k}}$. This is a linear
combination of $d$ orthonormal operators $\ket{\psi_{k}}\bra{\psi_{k}}$ with
coefficients corresponding to the eigenvalues of $M$. Consequently no orthonormal
basis of the space of operators with common eigenbasis to $M_a$ can have more than
$d$ elements and as such the maximal number of commuting, pairwise orthogonal unitaries 
$M_a$ is $d$. 

As a corollary, we obtain that the set 
$\tilde{\mathcal{M}}_{a}=\left\{ \left(M_{a}\right)^{b}\right\} _{b=1,\ldots,d-1}$
with the spectrum of all elements given by
Eq.~\eqref{eq:spectrum} together with the identity forms an Abelian
group of pairwise orthonormal operators with matrix multiplication as
group operation. $\tilde{\mathcal{M}}_{a}$ contains all the 
unitaries that share an eigenbasis with $M_{a}$ while having the
same spectrum as $M_{a}$ and being pairwise orthogonal.

\subsection{Complete set of measurement operators}
\label{subsec:complete}

The complete set of measurement operators $\tilde{\mathcal M}$ is obtained
iteratively by choosing a starting point, i.e., an operator $M_a$ with
spectrum according to Eq.~\eqref{eq:spectrum}. $M_a$ defines the
commmuting set $\tilde{\mathcal{M}}_a$ with all operators in
$\tilde{\mathcal{M}}_a$ given by Eq.~\eqref{eq:M_b}. Next one needs to
find another matrix $M_{a^\prime}$ with the same spectrum,
Eq.~\eqref{eq:spectrum}, but orthogonal to all
$M_a\in\tilde{\mathcal{M}}_a$. By construction, $M_{a^\prime}$ does
not share an eigenbasis with the
$M_a\in\tilde{\mathcal{M}}_a$. Rather, it defines, according to
Eq.~\eqref{eq:M_b}, its own set of commuting operators,
$\tilde{\mathcal{M}}_{a^\prime}$ which, together with the identity,
forms another Abelian group. The last step needs to be repeated until
$d+1$ Abelian groups $\tilde{\mathcal{M}}_{a}\cup\openone$ have been
found. The procedure of identifying $d+1$ sets of $d$ commuting,
pairwise orthogonal measurement operators yields, without
double-counting the identity which is an element of all the Abelian
groups, $d^2$ orthogonal measurement operators, i.e., the complete
operator basis $\mathcal M$. 

Clearly, one cannot find more than $d+1$ Abelian groups of orthogonal
operators since there exist only $d^{2}$ orthogonal operators on a
$d$-dimensional Hilbert space. Note that we know of the existence of
at least one such set of Abelian groups -- the generalized Pauli
operator basis $\mathcal{P}$ and its separation into mutually commuting subsets.
The operators belonging to the generalized Pauli basis are given
by~\cite{GottesmanChaos99,BandyoAlgo02,LawrencePRA02,LawrencePRA04}
\begin{subequations}\label{eq:pauligroup}
\begin{equation}
  \label{eq:pauliop}
  X^aZ^b\,,\quad a, b\in[0,d-1]\,,
\end{equation}
where $\omega=\exp{(2i\pi/d)}$ and 
\begin{eqnarray}
  \label{eq:X}
  X &=&\ket{n\oplus1}\bra{n}\,,\\
  \label{eq:Z}
  Z &=& \omega^n\ket{n}\bra{n}\,,
\end{eqnarray}
with $n\in[0,d-1]$ and addition is modulo $d$.
\end{subequations}

\section{Mutually unbiased bases}
\label{sec:MUB}

The existence of $d+1$ Abelian groups $\tilde{\mathcal{M}}_a\cup\openone$ 
of orthogonal measurement
operators is in a one-to-one correspondence to the existence of $d+1$
mutually unbiased bases~\cite{BandyoAlgo02}. 
This is easily seen using our constructions of Sec.~\ref{sec:spec}
and~\ref{sec:ebases}: The common 
eigenbasis of $\tilde{\mathcal{M}}_a$,
$\{\ket{\psi_{k}^{a}}\}$, can be used to construct
an operator basis,
\[
M_{au}=\left(M_{a}\right)^u =\sum_{k}e^{i\frac{2\pi}{d} uk}
\ket{\psi_{k}^{a}}\bra{\psi_{k}^{a}}\,.
\]
Projectors can be defined in terms of the operator basis, 
that is, 
\[
P_{n}^{a}=\ket{\psi_{n}^{a}}\bra{\psi_{n}^{a}}=
\frac{1}{d}\sum_{u}e^{-i\frac{2\pi}{d}un}\left(M_{a}\right)^u\,.
\]
Then
\[
\left|\braket{\psi_{n}^{a}|\psi_{n'}^{b}}\right|^{2}
=\Tr\left[P_{n}^{a}\left(P_{n'}^{b}\right)^{\dagger}\right]
=\frac{1}{d^{2}}\sum_{uu'}e^{-i\frac{2\pi}{d}\left(un-u'n'\right)}
\Tr\left[M_{au}M_{bu'}^{\dagger}\right]
\]
If $M_{a}$ and $M_{b}$ are from different Abelian groups, 
only identity ($u=u'=0$) contributes due to orthogonality of all other
measurement operators. In this case
\begin{equation}
\label{eq:MUB}
\left|\braket{\psi_{n}^{a}|\psi_{n'}^{b}}\right|^{2}
=\frac{1}{d^{2}}\Tr\left[\openone\right]=\frac{1}{d}\,.
\end{equation}
If $M_{a}$ and $M_{b}$ are from the same set
$\tilde{\mathcal{M}}_a\cup\openone$, all $u=u'$ contribute
and then
\[
\left|\braket{\psi_{n}^{a}|\psi_{n'}^{a}}\right|^{2}
=\frac{1}{d^{2}}\sum_{u}e^{-i\frac{2\pi}{d}u\left(n-n'\right)}
\Tr\left[M_{au}M_{au}^{\dagger}\right]
=\frac{1}{d}\sum_{u}e^{-i\frac{2\pi}{d}u\left(n-n'\right)}=\delta_{nn'}\,.
\]

The identification of the eigenbases of the measurement operators with
mutually unbiased bases allows us to determine which unitaries can be 
efficiently characterized with this operator basis. The candidate
unitaries need to map any measurement operator onto
another measurement operator from the set, modulo a phase corresponding
to a $d$th root of unity. Consider a specific 
measurement operator $M$ from an optimal set
$\tilde{\mathcal{M}}^\star$. $M$ 
is mapped by the candidate unitaries either to the same or to a different
Abelian group in $\tilde{\mathcal{M}}^\star$.
Given the spectral decomposition of $M$ in terms of its
eigenbasis, $\{\ket{\psi_{k}^{a}}\}$, with eigenvalues 
$\lambda_a$, we can write
\[
UMU^\dagger = \sum_a \lambda_a U\ket{\psi_{k}^{a}}\bra{\psi_{k}^{a}}U^\dagger
= \sum_a \lambda_a \ket{U\psi_{k}^{a}}\bra{U\psi_{k}^{a}} \overset{!}{=} M'\,,
\]
where $M'\in\tilde{\mathcal{M}}^\star$ by definition of $U$. Since the
$\{\ket{\psi_{k}^{a}}\}$ are orthonormal, 
so are the $\{\ket{U\psi_{k}^{a}}\}$; hence they correspond to the
eigenbasis of $M'$. Consequently, the set $\{\ket{U\psi_{k}^{a}}\}$
must either be identical to the set $\{\ket{\psi_{k}^{a}}\}$ modulo
phasefactors on the individual states or correspond to a basis which
is mutually unbiased to $\{\ket{\psi_{k}^{a}}\}$. Therefore
a unitary $U$ is efficiently characterizable if and
only if it keeps the partitioning of
the $d+1$ mutually unbiased bases in a Hilbert space of prime
dimension $d$ intact.

\section{Tensor products}
\label{sec:tensor}

We now consider $N$ qudits ($N>1$) and assume the measurement
operators to be tensor products of single-qudit operators. This choice
is motivated by the requirement to allow for product input states
since the preparation of these states is experimentally much easier. 
Product input states imply a tensor product structure for the
measurement basis since, in Monte Carlo estimation of the average
fidelity, the input states are the eigenstates of the measurement
basis~\cite{FlammiaPRL11,daSilvaPRL11,Giulia14}. 

Assuming the  measurement basis to be given by tensor products, we
obtain a natural partition of the total Hilbert space into a tensor
product of smaller Hilbert spaces. It corresponds to the direct
product structure imposed on the measurement basis. 
A natural approach to identify optimal measurement bases on the
total Hilbert space starts from maximizing the number of efficiently
characterizable unitaries on each subspace~\cite{Giulia14}. 
This is achieved by finding an optimal measurement basis on each
subspace as discussed above, provided the dimension of the subspace is
prime. The optimal measurement basis of the
total Hilbert space is then constructed in terms of tensor products of
the operators defined on the  subspaces.  
This yields indeed an orthonormal basis of measurement operators
on the total Hilbert space.

The dimension of each subspace is prime for $N$ identical qudits but also for
mixtures of e.g. qubits and qutrits ($d=3$).
If a subspace has  non-prime dimension, we suggest to perform
a prime decomposition of the dimension and construct the measurement
basis as tensor products of the optimal bases defined on the resulting prime
dimension subspaces,  analogously to the discussion above. Most
likely, this yields an optimal measurement basis. However, 
it remains an open question whether the explicit use of non-prime dimension
subspaces can be used to increase 
the number of efficiently characterizable unitaries beyond the one following
from the prime factor decomposition approach. Nonetheless, our
conjecture that a measurement basis constructed from the prime factor
decomposition represents indeed an optimal choice is motivated by the
fact that existence of $d+1$ mutually unbiased bases is not guaranteed
for non-prime dimension Hilbert spaces but seems to be a central
prerequisiste for obtaining efficiently characterizable unitaries~\cite{Giulia14}. 

\section{Conclusions}
\label{sec:concl}

Efficient estimation of the average fidelity of Clifford gates relies
on the property of these unitaries to map the basis of measurement
operators onto itself, up to a phase factor. We have used this
property to define optimality of a measurement basis in terms of the
maximum number of unitaries that can be efficiently characterized. For
Hilbert spaces of prime dimension, we have shown that this definition
yields a constructive proof for the optimal measurement basis and also
allows for identifying the unitaries which can be efficiently
characterized. For $N$ identical qudits, an optimal measurement basis
is obtained in terms of tensor products of the single-qudit operators
making up the optimal single-qudit operator basis. This choice
guarantees that the measurements are local in the sense that only
separable input states are required. 

Our construction of an optimal set of measurement operators with
the corresponding set of measurement bases
is determined only up to a global rotation. In other words, the choice
of the eigenbasis for the first Abelian group of measurement
operators is arbitrary. This corresponds to mutual unbiasedness
being defined only in relation of one basis to another.
If, in a given experimental setting, it is possible to perform
measurements and prepare input states relative to a rotated set of
mutually unbiased bases, this can be used to also rotate the set of efficiently
characterizable unitaries. Specifically, for any unitary $U$ there
exists a measurement basis in which $U$ can be efficiently
characterized. 
This is essentially the idea underlying randomized
benchmarking~\cite{MagesanPRL11} where arbitrary unitaries are rotated
into identity. The corresponding rotation on the input states
requires, however, application of the inverse of the unitary that
shall be characterized. This is  in general not practical.
In other words, the freedom of choice  for the global rotation of the
measurement can in principle be used to tune the set of efficiently
characterizable unitaries. Typically, however, the choice of the
eigenbasis  for the first Abelian group of measurement
operators is dictated by experimental convenience such as the
requirement of a separable eigenbasis. This fixes the set of unitaries that
can be characterized efficiently. 

The fact that our proof relies on the dimension of the Hilbert
(sub)spaces to be prime highlights the intimate relation between
finding efficiently characterizable unitaries and the existence of
mutually unbiased bases. In particular, for prime dimensions 
we have proven that the optimal basis of measurement operators can be
partitioned into $d+1$ commuting sets, i.e., it gives rise to 
a maximal partitioning. The generalized Pauli
operators~\cite{GottesmanChaos99,BandyoAlgo02,LawrencePRA02,LawrencePRA04}
represent one 
example of such an optimal measurement basis. Generalized Pauli
operators can also be defined for Hilbert spaces whose dimension cannot
be expressed as $d^N$ with $d$ prime~\cite{BermejoQuantInf14}. It
would be interesting to see whether in this case mutually unbiased
bases can be determined from the properties of the generalized Pauli
operators. 

\begin{acknowledgments}
  GG acknowledges support from a MIUR-PRIN grant (2010LLKJBX). QSTAR
  is the MPQ, LENS, IIT, 
  UniFi Joint Center for Quantum Science and Technology in Arcetri.
\end{acknowledgments}

\appendix 

\section{Details of the proofs}
\label{sec:app}

\subsection{The solution to
  Eq.~\eqref{eq:unityroots} is $c_s=1$} 
\label{subsec:unityroots}

We show here that the only solution of Eq.~\eqref{eq:unityroots}
for $c_s=0,1,\ldots \in \mathbb{N}$ under the additional constraint 
\begin{equation}
  \sum_{s=0}^{d-1}c_{s}=d
\end{equation}
is 
\[
c_s = 1
\]
for all $s$.
To prove this we use the fact that $e^{i\frac{2\pi s}{d}}$ is a 
$d$th root of unity for all $s$ and then apply a theorem of
Ref.~\cite{LamJA00} about sums over roots of unity.
Abbreviating $e^{i\frac{2\pi}{d}}=\omega$, 
Eq.~\eqref{eq:unityroots} becomes
\begin{equation}
  \label{eq:sumunityroots2}
  \sum_{s=0}^{d-1}c_{s}\omega^s=0\,.
\end{equation}
Since all $c_{s}$ are non-negative integers, this can be rewritten as 
\begin{equation}
  \label{eq:sumunityroots3}
  \sum_{t=0}^{d-1}\Omega_t=0\,,
\end{equation}
where $\Omega_t$ is a d-th root of unity. 
We absorbed the integer values of $c_{s}$
into the $\Omega_t$ by allowing for repetitions in the sum. So for example
if a $c_{s}$ was greater than one, there would be multiple indices $t$ in
Eq.~\eqref{eq:sumunityroots3} with $\Omega_t = \omega^s$. Furthermore, some
$c_{s}$ could be zero which means that the corresponding root of unity
$\omega^s$ does not appear in the set of $\Omega_t$.
Note furthermore that since we know
that the $c_s$ sum up to $d$, the sum in
Eq.~\eqref{eq:sumunityroots3} indeed has $d-1$ elements.

Consider now the general situation of sums over $d$th roots of
unity with an arbitrary number of summands, $n$. As in
Eq.~\eqref{eq:sumunityroots3}, the same root of unity may appear
multiple times. Lam and Leung showed~\cite{LamJA00} that 
if $d$ is prime, such a sum can only be equal to zero if $n$ is equal
to a multiple of $d$.  
As a consequence there exists no proper subsum of the sum in
Eq.~\eqref{eq:sumunityroots3} that goes to zero by itself. This
property is called minimal. Moreover, corollary 3.4. from
Ref.~\cite{LamJA00} implies that for $d$ prime the only minimal 
vanishing sum of $d$ roots of unity, including repetitions, is given by
\begin{equation}
  \sum_{t=0}^{d-1}\omega^t=e^{i\frac{2\pi t}{d}}=0\,.
\end{equation}
This translates into the sum in Eq.~\eqref{eq:sumunityroots3} 
having no repetitions but every root of unity appears exactly once.
Consequently, $c_s = 1$ in for all $s$ in
Eq.~\eqref{eq:unityroots} and the statement is proven.

\subsection{All matrices constructed according to
  Eq.~\eqref{eq:U_construction} are unitary for $d$ prime}
\label{subsec:unitary}

We show here that all matrices constructed according to
Eq.~\eqref{eq:U_construction} are unitary for $d$ prime
and contain on each (secondary) diagonal  only one entry equal to one. 

For simplicity we use normal addition symbols in this section but 
all algebraic manipulations are to be understood modulo $d$.
We first that each (secondary) diagonal contains only one entry equal
to one with all others being zero. Consider a
fixed $b$ and a fixed diagonal $t$. An element on this (secondary) 
diagonal, $(U^{ab}_{i,i+t})$ with $i=1,\ldots,d$, is nonzero  according to
Eq.~\eqref{eq:U_construction} if and
only if $\delta_{i+t,(i-1) \cdot b + 1} = 1$. To prove that,  
given $b$ and $t$, there is exactly one $i$ for which this can happen,
we consider the solutions of the equation
\begin{equation}
  \label{eq:diagcond} 
  \left(i-1\right)\cdot\left(b-1\right) = t\,,
\end{equation}
which follows directly from $\left(i-1\right)\cdot b+1 = i + t$.
If we keep $b$ fixed, showing that for each $t$
there is one $i$ for which Eq.~\eqref{eq:diagcond}
is fulfilled is equivalent to showing that for each $i$ there is exactly
one $t$ for which Eq.~\eqref{eq:diagcond} is fulfilled, i.e. $t(i)$ is
bijective. Then in each row a different diagonal acquires
the value $1$. Since the map $t(i)$ maps the finite set ${1,\dots,d}$
onto itself, injectivity implies surjectivity. Hence we only need to
prove that $t(i)$ is injective.

To do this, 
we need to find out how many solutions $i$ are allowed for
Eq.~\eqref{eq:diagcond} 
with $t\in\left\{ 0,\dots,d-1\right\}$. 
At least one solution to Eq.~\eqref{eq:diagcond} must exist
since by construction of $U_{ab}$, there is one entry equal to $1$ on
each row. According to the rules of modulo algebra, if one solution exists, 
then there are $g$ solutions with $g=\gcd\left(b-1,d\right)$ where $\gcd$
denoting the greatest common divisor. Since $b<d$
and $d$ is a prime number, $g=1$ and there exists only one
solution. This proves injectivity of $t(i)$.
Therefore the map $t(i)$
is bijective and so is $i(t)$ which implies 
that the construction of $U_{ab}$, Eq.~\eqref{eq:U_construction}, 
indeed fulfills the condition of exactly one entry equal to 1 on each
(secondary) diagonal. 

Next we show unitarity of $U_{ab}$. By construction, there exists
exactly one entry equal to $1$ in each row. It remains to be
shown that in each column there exists also only one 
entry equal to $1$. Unitarity of $U_{ab}$ then follows immediately. 

Let us consider for fixed $b$ a column $k$. According to
Eq.~\eqref{eq:U_construction}, an entry in the $i$th
row is nonzero if and only if $\delta_{k,(i-1) \cdot b + 1} = 1$. To
show that for fixed $b$ and $k$ there is exactly one $i$ for 
which this can happen, we consider the solutions of the equation
\begin{equation}
  \label{eq:columncond}
  (i-1) \cdot b + 1 = k\,.  
\end{equation}
Equation~\eqref{eq:columncond} defines a map $k(i)$. Showing that $k(i)$ is
bijective implies that for each $k$ 
there exists only one $i$ as a solution and vice versa, i.e., 
for each column there is only one row with an entry equal to
$1$. Employing the same argument as above,
there exist $g$ solutions to Eq.~\eqref{eq:columncond}
with $g=\gcd\left(b,d\right)$ and, since $b<d$ and $d$ is prime, $g=1$
and there exists only one solution. 
As a consequence the map $k(i)$ is bijective and so is the map $i(k)$,
i.e., the $U_{ab}$ constructed according to
Eq.~\eqref{eq:U_construction} are indeed unitary.


\begin{thebibliography}{13}
\expandafter\ifx\csname natexlab\endcsname\relax\def\natexlab#1{#1}\fi
\expandafter\ifx\csname bibnamefont\endcsname\relax
  \def\bibnamefont#1{#1}\fi
\expandafter\ifx\csname bibfnamefont\endcsname\relax
  \def\bibfnamefont#1{#1}\fi
\expandafter\ifx\csname citenamefont\endcsname\relax
  \def\citenamefont#1{#1}\fi
\expandafter\ifx\csname url\endcsname\relax
  \def\url#1{\texttt{#1}}\fi
\expandafter\ifx\csname urlprefix\endcsname\relax\def\urlprefix{URL }\fi
\providecommand{\bibinfo}[2]{#2}
\providecommand{\eprint}[2][]{\url{#2}}

\bibitem[{\citenamefont{Nielsen and Chuang}(2000)}]{NielsenChuang}
\bibinfo{author}{\bibfnamefont{M.~A.} \bibnamefont{Nielsen}} \bibnamefont{and}
  \bibinfo{author}{\bibfnamefont{I.~L.} \bibnamefont{Chuang}},
  \emph{\bibinfo{title}{Quantum Computation and Quantum Information}}
  (\bibinfo{publisher}{Cambridge University Press}, \bibinfo{year}{2000}).

\bibitem[{\citenamefont{Gottesman}(1999)}]{GottesmanChaos99}
\bibinfo{author}{\bibfnamefont{D.}~\bibnamefont{Gottesman}},
  \bibinfo{journal}{Chaos, Solitons \& Fractals} \textbf{\bibinfo{volume}{10}},
  \bibinfo{pages}{1749 } (\bibinfo{year}{1999}).

\bibitem[{\citenamefont{Boykin et~al.}(2000)\citenamefont{Boykin, Mor, Pulver,
  Roychowdhury, and Vatan}}]{Boykin00}
\bibinfo{author}{\bibfnamefont{P.}~\bibnamefont{Boykin}},
  \bibinfo{author}{\bibfnamefont{T.}~\bibnamefont{Mor}},
  \bibinfo{author}{\bibfnamefont{M.}~\bibnamefont{Pulver}},
  \bibinfo{author}{\bibfnamefont{V.}~\bibnamefont{Roychowdhury}},
  \bibnamefont{and} \bibinfo{author}{\bibfnamefont{F.}~\bibnamefont{Vatan}},
  \bibinfo{journal}{Information Processing Letters}
  \textbf{\bibinfo{volume}{75}}, \bibinfo{pages}{101 } (\bibinfo{year}{2000}),
  ISSN \bibinfo{issn}{0020-0190}.

\bibitem[{\citenamefont{Flammia and Liu}(2011)}]{FlammiaPRL11}
\bibinfo{author}{\bibfnamefont{S.~T.} \bibnamefont{Flammia}} \bibnamefont{and}
  \bibinfo{author}{\bibfnamefont{Y.-K.} \bibnamefont{Liu}},
  \bibinfo{journal}{Phys. Rev. Lett.} \textbf{\bibinfo{volume}{106}},
  \bibinfo{pages}{230501} (\bibinfo{year}{2011}).

\bibitem[{\citenamefont{da~Silva et~al.}(2011)\citenamefont{da~Silva,
  Landon-Cardinal, and Poulin}}]{daSilvaPRL11}
\bibinfo{author}{\bibfnamefont{M.~P.} \bibnamefont{da~Silva}},
  \bibinfo{author}{\bibfnamefont{O.}~\bibnamefont{Landon-Cardinal}},
  \bibnamefont{and} \bibinfo{author}{\bibfnamefont{D.}~\bibnamefont{Poulin}},
  \bibinfo{journal}{Phys. Rev. Lett.} \textbf{\bibinfo{volume}{107}},
  \bibinfo{pages}{210404} (\bibinfo{year}{2011}).

\bibitem[{\citenamefont{Magesan et~al.}(2011)\citenamefont{Magesan, Gambetta,
  and Emerson}}]{MagesanPRL11}
\bibinfo{author}{\bibfnamefont{E.}~\bibnamefont{Magesan}},
  \bibinfo{author}{\bibfnamefont{J.~M.} \bibnamefont{Gambetta}},
  \bibnamefont{and} \bibinfo{author}{\bibfnamefont{J.}~\bibnamefont{Emerson}},
  \bibinfo{journal}{Phys. Rev. Lett.} \textbf{\bibinfo{volume}{106}},
  \bibinfo{pages}{180504} (\bibinfo{year}{2011}).

\bibitem[{\citenamefont{Gualdi et~al.}(2014)\citenamefont{Gualdi, Licht, Reich,
  and Koch}}]{Giulia14}
\bibinfo{author}{\bibfnamefont{G.}~\bibnamefont{Gualdi}},
  \bibinfo{author}{\bibfnamefont{D.}~\bibnamefont{Licht}},
  \bibinfo{author}{\bibfnamefont{D.~M.} \bibnamefont{Reich}}, \bibnamefont{and}
  \bibinfo{author}{\bibfnamefont{C.~P.} \bibnamefont{Koch}},
  \bibinfo{journal}{arXiv:1404.1608}  (\bibinfo{year}{2014}).

\bibitem[{\citenamefont{Bermejo-Vega and Van
  Den~Nest}(2014)}]{BermejoQuantInf14}
\bibinfo{author}{\bibfnamefont{J.}~\bibnamefont{Bermejo-Vega}}
  \bibnamefont{and} \bibinfo{author}{\bibfnamefont{M.}~\bibnamefont{Van
  Den~Nest}}, \bibinfo{journal}{Quant. Info. Comput.}
  \textbf{\bibinfo{volume}{14}}, \bibinfo{pages}{0181} (\bibinfo{year}{2014}).

\bibitem[{\citenamefont{Wootters and Fields}(1989)}]{WoottersAnnPhys89}
\bibinfo{author}{\bibfnamefont{W.~K.} \bibnamefont{Wootters}} \bibnamefont{and}
  \bibinfo{author}{\bibfnamefont{B.~D.} \bibnamefont{Fields}},
  \bibinfo{journal}{Ann. Phys.} \textbf{\bibinfo{volume}{191}},
  \bibinfo{pages}{363} (\bibinfo{year}{1989}).

\bibitem[{\citenamefont{Bandyopadhayay
  et~al.}(2002)\citenamefont{Bandyopadhayay, Boykin, and
  Roychowdhury}}]{BandyoAlgo02}
\bibinfo{author}{\bibfnamefont{S.}~\bibnamefont{Bandyopadhayay}},
  \bibinfo{author}{\bibfnamefont{P.~O.} \bibnamefont{Boykin}},
  \bibnamefont{and} \bibinfo{author}{\bibfnamefont{V.~V.~F.}
  \bibnamefont{Roychowdhury}}, \bibinfo{journal}{Algorithmica}
  \textbf{\bibinfo{volume}{34}}, \bibinfo{pages}{512} (\bibinfo{year}{2002}).

\bibitem[{\citenamefont{Lawrence}(2004)}]{LawrencePRA04}
\bibinfo{author}{\bibfnamefont{J.}~\bibnamefont{Lawrence}},
  \bibinfo{journal}{Phys. Rev. A} \textbf{\bibinfo{volume}{70}},
  \bibinfo{pages}{012302} (\bibinfo{year}{2004}).

\bibitem[{\citenamefont{Lawrence et~al.}(2002)\citenamefont{Lawrence, Brukner,
  and Zeilinger}}]{LawrencePRA02}
\bibinfo{author}{\bibfnamefont{J.}~\bibnamefont{Lawrence}},
  \bibinfo{author}{\bibfnamefont{C.}~\bibnamefont{Brukner}}, \bibnamefont{and}
  \bibinfo{author}{\bibfnamefont{A.}~\bibnamefont{Zeilinger}},
  \bibinfo{journal}{Phys. Rev. A} \textbf{\bibinfo{volume}{65}},
  \bibinfo{pages}{032320} (\bibinfo{year}{2002}).

\bibitem[{\citenamefont{Lam and Leung}(2000)}]{LamJA00}
\bibinfo{author}{\bibfnamefont{T.}~\bibnamefont{Lam}} \bibnamefont{and}
  \bibinfo{author}{\bibfnamefont{K.}~\bibnamefont{Leung}}, \bibinfo{journal}{J.
  Algebra} \textbf{\bibinfo{volume}{224}}, \bibinfo{pages}{91 }
  (\bibinfo{year}{2000}).

\end{thebibliography}

\end{document}